\documentclass[aps,pre,reprint,superscriptaddress,amsmath,amssymb,showpacs,floatfix,longbibliography]{revtex4-1}
\usepackage{graphicx}
\usepackage{bm}
\usepackage{color}

\def\dbar{{\mathchar'26\mkern-12mu d}}

\begin{document}

\title{Hidden entropy production by fast variables} 
\author{Hyun-Myung Chun}
\affiliation{Department of Physics, University of Seoul, Seoul 130-743,
 Korea}
\author{Jae Dong Noh}
\affiliation{Department of Physics, University of Seoul, Seoul 130-743,
 Korea}
\affiliation{School of Physics, Korea Institute for Advanced Study,
Seoul 130-722,  Korea}

\date{April 7, 2013}

\begin{abstract}
We investigate nonequilibrium underdamped Langevin dynamics of 
Brownian particles that interact through a harmonic potential with coupling
constant $K$ and are in thermal contact with two heat baths at different
temperatures. The system is characterized by a net heat flow and an entropy
production in the steady state. We compare the entropy production of the
harmonic system with that of Brownian particles linked with a rigid rod.
The harmonic system may be expected to reduce to the rigid rod system in the 
infinite $K$ limit. 
However, we find that the harmonic system in the $K\to\infty$ limit produces 
more entropy than the rigid rod system. The harmonic system has the center
of mass coordinate as a slow variable and the relative coordinate as a fast
variable. By identifying the contributions of the degrees of
freedom to the total entropy production, we show that 
the hidden entropy production by the fast variable is responsible for the
extra entropy production. We discuss the $K$ dependence of each contribution.

\end{abstract}
\pacs{05.70.Ln, 05.40.-a, 02.50.-r, 05.10.Gg}
\maketitle

\section{Introduction}
Recent developments in stochastic thermodynamics boost active researches 
on microscopic systems in thermal environments.
Stochastic thermodynamics allows one to define thermodynamic quantities 
such as heat and work at the microscopic trajectory level, 
with which thermodynamic properties are readily 
studied~\cite{Seifert:2005fu,Sekimoto:1998uf}.

When one studies a thermodynamic system, a coarse-grained description may be  
necessary~\cite{Zamponi:2005dk,Rahav:2007el,
Pigolotti:2008go,Puglisi:2010dn,Santillan:2011gu,Esposito:2012jt,Mehl:2012fw,
Kawaguchi:2013ca,Baek:2014ca,Shiraishi:2015do}. 
Suppose that some transition rates among microscopic configurations are
much faster than the others. Such systems can be coarse-grained in terms of 
so-called mesostates which refer to a set of microscopic configurations 
connected via fast processes~\cite{Rahav:2007el,Esposito:2012jt}. The
coarse-graining is also necessary if some degrees of freedom are 
inaccessible due to e.g., a resolution limit of measurement
devices~\cite{Mehl:2012fw, Shiraishi:2015do}. 
The coarse-graining naturally raises questions on the formulation of 
effective thermodynamics and on the extent to which the thermodynamic laws 
are valid.

\begin{figure}
\includegraphics[width=\columnwidth]{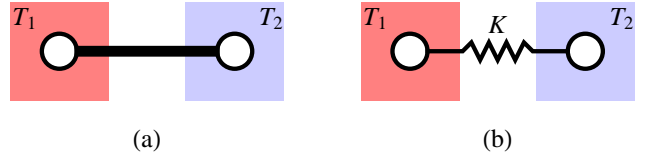}
\caption{Comparison of a rigid rod system in (a) and a harmonic system in
(b).}\label{fig1}
\end{figure}

In this paper, we investigate the entropy production by a harmonic chain of 
Brownian particles between two heat baths at different temperatures with
the focus on the the role of fast and slow degrees of freedom. When two heat
baths are connected with a medium, heat flows from a higher temperature
bath to a lower temperature one and the total entropy
increases~\cite{{Visco:2006en},{Kundu:2011hd},{Fogedby:2011it}}. 
A toy model for this phenomenon in one dimension 
was studied thoroughly in
Refs.~\cite{Visco:2006en,{VandenBroeck:2001un}} with
the equation of motion for the displacement $X$:
\begin{equation}\label{rod_eq}
M \frac{d^2X}{dt^2} = -\gamma_1 \frac{dX}{dt} + \xi_1(t) 
                      -\gamma_2 \frac{dX}{dt} + \xi_2(t) \ ,
\end{equation} 
where $\gamma_i~(i=1,2)$ is the damping coefficient of a heat bath $i$ at
temperature $T_i$ and $\xi_i(t)$ is the thermal Langevin noise satisfying 
\begin{equation}\label{noise_corr}
\langle \xi_i(t) \rangle = 0, \ \langle \xi_i(t) \xi_j(t')\rangle =
2\gamma_i T_i \delta_{ij} \delta(t-t') \ .
\end{equation}
The Boltzmann constant $k_B$ is set to unity.
The model describes a single Brownian particle of mass $M$ coupled to two
heat baths simultaneously. At the same time, as illustrated in 
Fig.~\ref{fig1}(a), it also describes two Brownian particles of total mass $M$ 
tied with a rigid rod. This rigid rod system was also studied in
Refs.~\cite{{VandenBroeck:2001un},{Parrondo:1996vh},{GomezMarin:2006jf}} for
the heat conduction in the Feynman ratchet~\cite{Feynman:1963tm}.

We notice that the model with a single degree of
freedom assumes an ideal medium for heat conduction, which could be  
realized with a many-body system in the strong coupling limit. 
A many-body system has multiple degrees of freedom such as the center 
of mass coordinate and the relative coordinates. In the infinite coupling
limit, the relative coordinates relaxing fast can be coarse-grained out
to leave the center of mass coordinate as a relevant slow variable.
It is interesting to investigate the similarity and the 
dissimilarity between the system given by \eqref{rod_eq} and a system with
multiple degrees of freedom in the strong coupling limit. 

For the purpose, we consider two Brownian particles exchanging a 
harmonic interaction as shown in Fig.~\ref{fig1}(b). 
Heat conduction and the entropy production in harmonic systems have long
been studied~\cite{Rieder:1967fi,Bernardin:2012fd, Dhar:2011ta,
Sabhapandit:2012vc,Saito:2011et,Kundu:2011hd,Fogedby:2012ku}.
In contrast to the previous works, our study focuses on the role of the slow 
and fast variable. 
Harmonic systems are analytically tractable~\cite{Kwon:2011fk,
Kwon:2013cl, Noh:2014if, Noh:2012hg,{Noh:2013jm},{Ciliberto:2004ei}}. 
We derive the formula for the entropy productions by the slow and fast
variables as a function of the coupling constant, which will shed a light on
the effect of the coarse graining.

This paper is organized as follows: The model system consisting of two
Brownian particles is introduced in Sec.~\ref{sec:2}. 
The heat flow and the entropy production in this model are calculated 
in Sec.~\ref{sec:3}. The slow
and fast variable are introduced and their contributions to the total
entropy production are investigated in Sec.~\ref{sec:4}. 
We compare the harmonic system and the rigid rod system, and draw the
conclusion for the hidden entropy production by the fast variable. We
conclude the paper with summary and discussions in Sec.~\ref{sec:5}.

\section{Two particle system}\label{sec:2}
Two particles of mass $m$ in one dimension
interact through a harmonic potential 
\begin{equation}\label{V12}
V(x_1,x_2) = \frac{K}{2}(x_1-x_2)^2 , 
\end{equation} 
where $x_i$~($i=1,2$) is the displacement of a particle $i$ and 
$K$ is the coupling constant.
A particle $i$ is in contact with the $i$th heat bath which is 
characterized with a temperature $T_i$ and a damping coefficient
$\gamma_i$~(see Fig.~\ref{fig1}(b)). 
The underdamped Langevin equation is given by 
\begin{equation}\label{Langevin_eq_N2}
\begin{split}
\dot{x}_1 &= v_1 , \\
\dot{x}_2 &= v_2 , \\
m \dot{v}_1 &= -\gamma_1 {v}_1 - K(x_1-x_2) + \xi_1(t) , \\ 
m \dot{v}_2 &= -\gamma_2 {v}_2 - K(x_2-x_1) + \xi_2(t) ,
\end{split}
\end{equation}
where $\{\xi_i(t)\}$ represent the thermal noises satisfying \eqref{noise_corr}.

Suppose that the system evolves from a configuration
$(x_1,v_1;x_2,v_2)$ at time $t$ to $(x_1+dx_1,v_1+dv_1;x_2+dx_2,v_2+dv_2)$ 
in an infinitesimal time step $dt$. 
According to stochastic thermodynamics~\cite{Seifert:2005fu}, the total entropy
production is given by
\begin{equation}
dS_{tot} = dS_s + dS_{b} , \
\end{equation}
where $dS_{s}$ denotes the Shannon entropy change of the Brownian
particles and $dS_{b}$ is the entropy change of the thermal heat baths. The
latter is given by 
\begin{equation}
dS_b = -\frac{\dbar Q_1}{T_1} - \frac{\dbar Q_2}{T_2}
\end{equation}
where
\begin{equation}\label{dQ_def}
\dbar Q_i = v_i \circ \left( -\gamma_i v_i dt + dW_i \right)
\end{equation}
is the heat absorbed by the Brownian particle from the bath $i$, and
$dW_i \equiv \int_t^{d+dt}dt' \xi_i(t')dt'$ satisfies $\langle
dW_i\rangle = 0$ and $\langle (dW_i)^2\rangle = 2\gamma_i T_i dt$.
The notation $\circ$ means that the product is in the Stratonovich
sense~\cite{{Gardiner:2010tp}}.

The heat can be expressed in different ways. 
By replacing $(-\gamma_1 v_1 dt + dW_1)$ with $(m dv_1 + K(x_1-x_2)dt)$, 
we obtain
\begin{equation}\label{dQ1}
\begin{split}
\dbar Q_1 =& d\left( \frac{1}{2}mv_1^2 + \frac{1}{2}K x_1^2 -
\frac{1}{2}Kx_1x_2 \right) \\
& + \frac{K}{2} (x_1 v_2 - x_2 v_1) dt  .
\end{split} \end{equation}
Similarly, we also obtain 
\begin{equation}\label{dQ2}\begin{split}
\dbar Q_2 =& d\left( \frac{1}{2}mv_2^2 + \frac{1}{2}K x_2^2 -
\frac{1}{2}Kx_1x_2 \right)  \\
& - \frac{K}{2} (x_1 v_2 -v_1 x_2) dt .
\end{split}\end{equation}
They add up to $dE = \dbar Q_1 + \dbar Q_2$, where $dE$ is the change in the
the total internal energy $E=\frac{1}{2}m(v_1^2+v_2^2) + V(x_1,x_2)$. 
The total derivatives in \eqref{dQ1} and \eqref{dQ2} 
do not contribute to the steady state average. 
Hence, the steady-state average of the heat flux, denoted by $q_i \equiv 
\langle \dbar Q_i/dt\rangle_s$ is given by
\begin{equation}\label{q1q2}
q = q_1 = - q_2 = \frac{K}{2} \langle x_1 v_2 - x_2 v_1\rangle_s \ ,
\end{equation}
where $\langle \cdot \rangle_s$ means the steady state ensemble average.

The heat flux acquires an interesting interpretation by
regarding $x_i$ and $v_i$ as the coordinates of the position vector $\bm{x} =
(x_1,x_2,0)$ and the velocity vector $\bm{v}=(v_1,v_2,0)$ 
of an imaginary particle moving on a $xy$ plane
in a three dimensional space. Then the heat flux in \eqref{q1q2} 
becomes proportional to the $z$ component of the angular momentum 
$\bm{L} = m \bm{x}\times \bm{v}$ of the imaginary particle: 
\begin{equation}
q = \frac{K}{2m} \langle L_3 \rangle_s \ .
\end{equation}
A nonzero heat flow implies a rotational motion. 
Such a rotational motion induced by a nonequilibrium 
driving was investigated in the context
of a Brownian gyrator~\cite{Filliger:2007fy}.

We find another useful expression for the heat flow.
The Stratonovich product $v_1 \circ dW_1$ in \eqref{dQ_def}
is defined as $\frac{1}{2}[v_1+(v_1+dv_1) ]
dW_1$~\cite{Gardiner:2010tp}, where 
$dv_1 = [-\gamma_1 v_1 dt - K(x_1-x_2) dt+dW_1]/m$ from the Langevin
equation. The ensemble average yields 
$\langle v_1 \circ dW_1\rangle = \frac{\gamma_1T_1}{m}dt$.
Thus, the average heat flux in the steady state can be written as
\begin{equation}\label{q1_simple}
q = \frac{2\gamma_1}{m}  \left( \frac{T_1}{2} - \frac{1}{2} m \langle
v_1^2\rangle_s \right) \ .
\end{equation}
The heat flow is driven by the imbalance between the temperature of 
the heat bath and the effective temperature of the particle measured 
by the kinetic energy. Note that the expression \eqref{q1_simple} is valid
not only in the steady state but also in a transient state.

In the steady state, the Shannon entropy of the particles does not
change~($\langle dS_s\rangle_s=0$).  
Consequently, the total entropy production rate in the steady state 
is given by
\begin{equation}
s \equiv \left\langle \frac{dS_{tot}}{dt}\right\rangle_s
  = \frac{-q}{T_1}+\frac{q}{T_2} = 
  \left(\frac{1}{T_2}-\frac{1}{T_1}\right) q \ .
\end{equation}
The heat flux and the entropy production rate is proportional to each other.

\section{Heat flow and entropy production}\label{sec:3}

For the heat flux, we need to evaluate the second
moments of the velocities, $M_{ij} \equiv \langle v_i(t) v_j(t)\rangle_s$.
The model with the linear Langevin equation in \eqref{Langevin_eq_N2}
belongs to the class of the multivariate Ornstein-Uhlenbeck 
process~\cite{Gardiner:2010tp}. 
The second moments of such a system satisfy a set of linear
equations~\cite{Gardiner:2010tp}, from which $M_{ij}$ can be easily
evaluated. Instead of using the method, we evaluate the moments explicitly
for self-containedness.

The Langevin equation \eqref{Langevin_eq_N2} is solved in the Fourier space. 
We introduce column vectors $\bm{\tilde{x}}(\omega) =
(\tilde{x}_1(\omega),\tilde{x}_2(\omega))^{\rm T}$ and 
$\bm{\tilde{\xi}}(\omega) =
(\tilde{\xi}_1(\omega),\tilde{\xi}_2(\omega))^{\rm T}$ with
$\tilde{x}_i(\omega) = \int_{-\infty}^\infty dt e^{i\omega t} x_i(t)$ and
$\tilde{\xi}_i(\omega) = \int_{-\infty}^\infty dt e^{i\omega t} \xi_i(t)$.
The superscript ${}^{\rm T}$ denotes the transpose. The Fourier components
of the thermal noise obey
\begin{equation}\label{xixi_w}
\left\langle \tilde{\xi}_i(\omega)\right\rangle = 0 \ , \
\left\langle \tilde{\xi}_i(\omega) \tilde{\xi}_j(\omega')\right\rangle 
= 4\pi \gamma_i T_i \delta_{ij} \delta(\omega+\omega') \ .
\end{equation}
The Langevin equation becomes 
$\mathsf{F}(\omega) \bm{\tilde{x}}(\omega) = \bm{\tilde{\xi}}(\omega)$,
where $\mathsf{F}(\omega)$ is a $2\times 2$ matrix given by
$$
\mathsf{F}(\omega) = \begin{pmatrix} 
   -m\omega^2-i\gamma_1\omega+K & -K \\
   -K & -m\omega^2-i\gamma_2\omega+K 
 \end{pmatrix}.
$$
The steady state solution is given by
$\bm{\tilde{x}}(\omega) = \mathsf{F}(\omega)^{-1} \bm{\xi}(\omega)$. 
The homogeneous solution is ignored because it vanishes
in the steady state.
For convenience, we write the solution as $\bm{\tilde{x}}(\omega) =
\frac{1}{D(\omega)}
\mathsf{C}(\omega) \bm{\tilde\xi}(\omega)$ where
\begin{equation}\label{denominator}
D(\omega) = (-m\omega^2 - i\gamma_1 \omega+K)
            (-m\omega^2-i\gamma_2 \omega+K)-K^2
\end{equation}
is the determinant of $\mathsf{F}(\omega)$ and
$\mathsf{C}(\omega)=\mathsf{C}(\omega)^{\rm T}$ is the
cofactor matrix given by
\begin{equation}\label{matrix_C}
\mathsf{C}(\omega) = \begin{pmatrix}
-m \omega^2-i\gamma_2 \omega + K & K  \\ 
K & -m \omega^2-i\gamma_1 \omega + K  
\end{pmatrix} .
\end{equation} 

By using $\tilde{v}_i(\omega) = -i \omega \tilde{x}_i(\omega)$ and the noise
correlations in \eqref{xixi_w}, we can write 
\begin{equation}\label{residue_int}
M_{ij} = \int_{-\infty}^{\infty} \frac{d\omega}{\pi}
\frac{\omega^2  [\sum_l \gamma_l T_l C_{il}(\omega)C_{jl}(-\omega)]}
{D(\omega)D(-\omega)} .
\end{equation}
This integral is evaluated using the contour integral method, 
as detailed in Appendix~\ref{app1}. 
The results are
\begin{equation}\label{M2_v}
\begin{split}
M_{11} &= 
   \frac{T_1}{m} - \frac{\gamma_2 K (T_1-T_2)}
   {(\gamma_1+\gamma_2)(Km+\gamma_1\gamma_2)}, \\
M_{22} &= 
   \frac{T_2}{m} + \frac{\gamma_1 K (T_1-T_2)}
   {(\gamma_1+\gamma_2)(Km+\gamma_1\gamma_2)}, \\
M_{12} &= M_{21} = 0 .
\end{split}
\end{equation}
Inserting these into \eqref{q1_simple}, we finally obtain 
\begin{equation}\label{q1_res}
q = \frac{\gamma_1\gamma_2 K
(T_1-T_2)}{(\gamma_1+\gamma_2)(Km+\gamma_1\gamma_2)} 
\end{equation}
for the heat flux and 
\begin{equation}\label{Sprod_res}
s = 
\frac{\gamma_1 \gamma_2 K}{(\gamma_1+\gamma_2) (Km+\gamma_1\gamma_2)} 
\frac{(T_1-T_2)^2}{T_1 T_2} 
\end{equation}
for the entropy production rate.

The heat flow is nonzero when $T_1\neq T_2$ and $K$ is nonzero. 
The system absorbs a heat from a higher temperature
heat bath and dissipates to a lower temperature one, 
which increases the entropy
of the whole system. 

\section{Hidden entropy production}\label{sec:4}
The heat flux and the entropy production rate increase monotonically with
the coupling constant $K$. In the $K\to\infty$ limit,
the heat flow rate converges to
\begin{equation}
q_\infty \equiv \lim_{K\to\infty} q =
\frac{\gamma_1\gamma_2}{m(\gamma_1+\gamma_2)}(T_1-T_2) ,
\end{equation}
and the entropy production rate converges to
\begin{equation}
s_\infty = \frac{\gamma_1\gamma_2}{m(\gamma_1+\gamma_2)}
\frac{(T_1-T_2)^2}{T_1T_2} \ .
\end{equation}
In this limit, the relative displacement of the two
particles~($|x_2-x_1|\sim K^{-1/2}$) vanishes.
Hence, one may regard the limiting case as the rigid rod system,
that is, a single Brownian particle of mass $M=2m$~(see Fig.~\ref{fig1}). 
It is worthy to compare the heat flux in both cases. 
The heat flux $q_{\rm rigid}$ of the rigid rod system 
is given by~Ref.~\cite{VandenBroeck:2001un,Visco:2006en}
\begin{equation}
q_{\rm rigid} = \frac{\gamma_1\gamma_2}{M(\gamma_1+\gamma_2)}(T_1-T_2) 
= \frac{\gamma_1\gamma_2}{2m(\gamma_1+\gamma_2)}(T_1-T_2) .
\end{equation}
Surprisingly, the heat flux $q_{\rm rigid}$ is the half of the limiting
values $q_\infty$ because of the factor $2$ in the denominator. 

In order to understand the origin of the discrepancy, we rewrite
the Langevin equation~\eqref{Langevin_eq_N2} in terms of the center of mass
coordinate $X=(x_1+x_2)/2$ and the relative coordinate $x=x_1-x_2$.
The equations of motion for $X$ and $x$ are given by 
\begin{equation}
\begin{split}
\dot{X} &=V, \\
\dot{x} &=v, \\
M \dot{V} &= - (\gamma_1+\gamma_2) {V} - \displaystyle 
   \frac{(\gamma_1-\gamma_2)}{2} v + \zeta_1(t), \\ 
   \mu \dot{v} &= \displaystyle -\frac{(\gamma_1+\gamma_2)}{4} {v} -
\frac{(\gamma_1-\gamma_2)}{2} V - K x + \zeta_2(t),
\end{split}
\end{equation}
where $M=2m$ is the total mass, $\mu = m/2$ is the reduced mass, 
$\zeta_1 \equiv \xi_1+\xi_2$ obeying 
$\langle \zeta_1(t)\rangle=0$ and $\langle \zeta_1(t)\zeta_1(t')\rangle =
2(\gamma_1 T_1+ \gamma_2 T_2)\delta(t-t')$, 
and $\zeta_2\equiv(\xi_1-\xi_2)/2$ obeying 
$\langle\zeta_2(t)\rangle=0$ and
$\langle\zeta_2(t)\zeta_2(t')\rangle =
\frac{1}{2}(\gamma_1 T_1+\gamma_2 T_2)\delta(t-t')$. 
The center of mass and the the relative coordinates are coupled 
through the noise correlation
$\langle \zeta_1(t)\zeta_2(t')\rangle = (\gamma_1 T_1-\gamma_2 T_2)
\delta(t-t')$ and the friction forces.

We can grasp an important feature of the system by considering 
the case with $\gamma_1=\gamma_2=\gamma$. 
The center of mass performs 
a Brownian motion with the effective temperature 
$T_{\rm eff} = (T_1+T_2)/2$, and the relative coordinate performs the
Ornstein-Uhlenbeck process~\cite{Gardiner:2010tp}
under the harmonic potential $\frac{1}{2}K x^2$ with 
the same effective temperature $T_{\rm eff}$.
The center of mass coordinate diffuse with the time scale $\tau_X \sim
m/\gamma$. 
The relative coordinate oscillates with the time scale $\tau_x \sim
\sqrt{m/K}$ with the amplitude $|x| \sim \sqrt{T_{\rm eff}/K}$. 
Therefore, in the $K\to\infty$ limit, $x$ becomes a {\em fast}
variable~($\tau_x \ll \tau_X$) and becomes frozen~($|x|\to 0$) in amplitude, 
while $X$ becomes a {\em slow} variable being governed by the same equation 
of motion in \eqref{rod_eq} as the rigid rod system.

Although the fast variable $x$ becomes frozen in the
$K\to\infty$ limit, it may leave a signature in thermodynamic 
quantities.
We scrutinize this possibility by decomposing the
heat flux, or equivalently the entropy production rate, into the sum of
contributions of the fast and the slow variables.
We start with the expression~\eqref{dQ1} for general $\gamma_1$ and
$\gamma_2$. After inserting $v_1 = V + v/2$ into \eqref{dQ1}, we obtain
\begin{equation}\begin{split}
\dbar Q_1 =& (-\gamma_1 V^2 dt + V\circ dW_1) + (-\gamma_1 v V dt )\\
     & + \left(-\frac{\gamma_1}{4} v^2 dt + \frac{v}{2}\circ dW_1\right) \ .
\end{split}\end{equation}
Each term corresponds to the heat flow via the slow variable, the
interplay between the slow and fast variables, and the fast variable,
respectively. Thus, the average heat flow rate in the steady state is 
decomposed as
\begin{equation}
q = q_S + q_{F} + q_m \ ,
\end{equation}
where
\begin{equation}\label{qS}
q_S = \frac{2\gamma_1}{M} \left( \frac{T_1}{2} - 
     \frac{1}{2} M \langle V^2\rangle_s\right) 
\end{equation}
is the contribution from the slow variable,
\begin{equation}\label{qF}
q_F = \frac{\gamma_1}{2\mu} \left( \frac{T_1}{2} - \frac{1}{2}\mu \langle
v^2\rangle_s \right)
\end{equation}
is from the fast variable, and
\begin{equation}\label{qm}
q_{m} = -\gamma_1 \langle vV\rangle_s
\end{equation}
is the mixed contribution. 

\begin{figure}
\includegraphics*[width=\columnwidth]{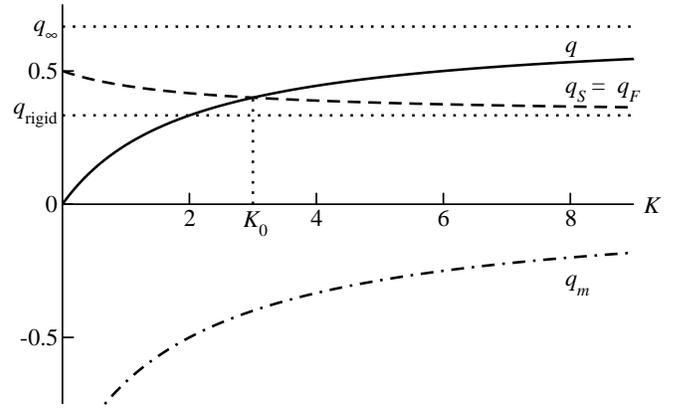}
\caption{Heat flow rates as a function of $K$. Parameter values
are $\gamma_1=2$, $\gamma_2=1$, $m=1$, and $(T_1-T_2)=1$.}
\label{fig2}
\end{figure}

Each contribution is easily evaluated using the second moments $M_{ij}$ 
obtained in \eqref{M2_v}. 
For example, $\langle V^2\rangle_s = (M_{11}+2M_{12}+M_{22})/4$.
The results are
\begin{eqnarray}
q_S &=& \frac{\gamma_1\gamma_2}{4m} \frac{2Km + \gamma_1^2+\gamma_1\gamma_2}
    {(\gamma_1+\gamma_2)(Km+\gamma_1\gamma_2)} (T_1-T_2) \\
q_{m} &=& - \frac{\gamma_1}{2m}
\frac{\gamma_1\gamma_2}{(Km+\gamma_1\gamma_2)} (T_1-T_2) \\
q_F &=& q_S \ .
\end{eqnarray}
Surprisingly, the fast variable contributes to the heat flow 
by the same amount as the slow variable at all values of $K$. 
This hidden entropy production by the fast variable explains the discrepancy
between $q_\infty$ and $q_{\rm rigid}$.
The mixed contribution $q_{m}$ has the opposite sign to $q_S$ and $q_F$, 
and vanishes in the infinite coupling limit.
They are plotted as a function of $K$ in Fig.~\ref{fig2}. 

Measurements are done within the resolution limit of probing devices. 
Suppose that one experiments on the harmonic system with large $K$ 
with a device whose time resolution $\tau_{\rm res} \gg \tau_x$. 
Then, the harmonic system would be observed as the rigid rod system.
Measurements would yield the apparent entropy production from the slow
variable only, which is a part of the total entropy production. 

The opposite case is also interesting. When the coupling constant 
vanishes~($K\to 0$), the two particles are in their own thermal
equilibrium and the heat flux and the entropy production vanish.
However, the apparent heat flux $q_S$ associated with the center of mass
coordinate still remains finite even at $K=0$.
Suppose that the two noninteracting particles are confined by an external
potential within a region whose size is much smaller than the spatial 
resolution of an experimental device. Then, the two particles would be 
considered as a single particle, which results in the overestimation of the
total entropy production.

The results are summarized in Fig.~\ref{fig2}. The heat flux $q$ increases
from $0$ to $q_\infty$ as $K$ increases from zero to infinity. 
The apparent heat flux $q_S$ is different from $q$. 
When $K$ is small, the hidden heat flux $(q_F + q_m)$ is negative 
so that the resolution limited measurement would overestimate the entropy 
production. When $K$ is large, the hidden heat flux
is positive so that the resolution limited measurement would underestimate 
the entropy production. The threshold $K_0$ is determined from the condition
$q=q_S$, which yields that
\begin{equation}
K_0 = \frac{\gamma_1 (\gamma_1+\gamma_2)}{2m} \ .
\end{equation}

The conclusion on the hidden entropy production does not depend on 
the specific form of the
interaction potential $V(x_1,x_2)$. We also considered the system with
$V(x_1,x_2) = \frac{K}{2}(x_1-x_2)^2+\frac{1}{2}k x_1^2+\frac{1}{2}k x_2^2$.
That is, each particle is trapped with the additional harmonic potential with
the coupling constant $k$. This case is also analyzed in Appendix~\ref{app1}. 
The total heat flux and the components are given in \eqref{kKcase}.
With this setting, we also obtain the same result that $q = 2 q_{\rm rigid}$ 
and that $q_S=q_F=q_{\rm rigid}$ with $q_m=0$ in the $K\to\infty$ limit.

\section{Summary and Discussions}\label{sec:5}
We have investigated the role of a fast variable in the entropy production. 
For the harmonic system, the total entropy production is
separated into the contributions from the slow variable~(center of mass
coordinate) and the fast variable~(relative coordinate), and the mixed
contribution. We found that the hidden entropy production due to the fast
variable remains finite even in the infinite coupling limit.

Our study reveals the difference between the rigid rod system and 
the harmonic system in the infinite coupling limit. In the former system,
the relative coordinate $x=x_1-x_2$ is constrained to be zero without any
dynamic fluctuations. In the latter system, the amplitude of the relative 
motion is also negligible~($x \sim \mathcal{O}(K^{-1/2})$). However, 
the velocity fluctuations are non-negligible, which results in a finite
additional hidden entropy production. 

The hidden entropy production could be relevant for heat engines. 
When one models a heat engine, some degrees of freedom may be ignored 
for the sake of simplicity.
For example, in the Feynman ratchet, vanes and a ratchet are assumed to be 
linked with a rigid
axle~\cite{Feynman:1963tm,{Parrondo:1996vh},{GomezMarin:2006jf}},
which ignore relative motions of the two parts. Our results suggest that the
entropy production in such a model should be less than the actual one
because of the hidden entropy.
The entropy production is the origin for the loss in the efficiency of 
a heat engine. Our result warns that all the fast variables should be taken
into account in analyzing the efficiency of a heat engine. 

\appendix
\section{Evaluation of the integral in \eqref{residue_int}}\label{app1}
\begin{figure}[t]
\includegraphics*[width=0.7\columnwidth]{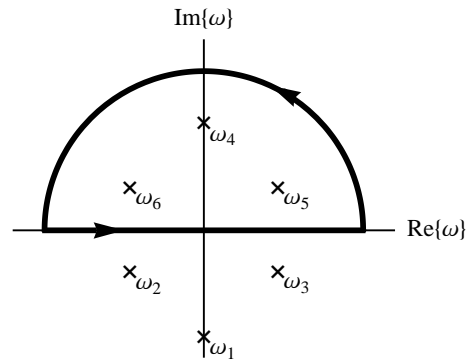}
\caption{Integration contour for the evaluation of \eqref{residue_int} 
in complex $\omega$ plane. Simple poles are marked with $\times$ symbols.
The non-zero roots $\omega_{4,5,6}$ of $D(-\omega)$ are 
in the upper half-plane, while the non-zero roots $\omega_{1,2,3}$
of $D(\omega)$ are in the lower half-plane.}
\label{fig3}
\end{figure}

The denominator~(numerator) of the integrand in \eqref{residue_int} is the
$8$th~(6th) degree polynomial in $\omega$. Thus, one can add a half-circular
contour of the infinite radius in the complex $\omega$ plane 
to make a closed contour as shown in Fig.~\ref{fig3}. 
The contour integral is determined by the residues of the poles
that are located at the roots of $D(\omega)$ and $D(-\omega)$ 
inside the contour. 

We first consider the roots of the quartic polynomial $D(\omega)$.
A trivial root is located at $\omega=0$. 
The others are the solutions of 
$g(\omega) \equiv m^2 \omega^3+i m(\gamma_1+\gamma_2)\omega^2- 
(\gamma_1 \gamma_2+2 m K)\omega - i(\gamma_1+\gamma_2) K=0$.
It is more convenient to work with 
$\tilde{g}(\Omega) \equiv i g(\omega = i\Omega) 
= m^2 \Omega^3 + m (\gamma_1+\gamma_2) 
\Omega^2 + (\gamma_1\gamma_2+2mK)\Omega+(\gamma_1+\gamma_2)K$ whose  
coefficients are real and positive. Positivity of coefficients
guarantees that real positive roots do not exist and that 
one of the roots, denoted by $\Omega_1$, must be real and negative. 
The remaining two roots, denoted by 
$\Omega_2$ and $\Omega_3$, may be real and negative, 
or a pair of complex conjugates. In order to determine whether they are
inside the contour in Fig.~\ref{fig3}, 
we need to know the sign of the real parts of $\Omega_{2}$ and $\Omega_3$.
The three roots of $\tilde{g}(\Omega)$ satisfy 
$\Omega_1+\Omega_2+\Omega_3 = -(\gamma_1+\gamma_2)/m$. 
Note that
$\tilde{g}(-(\gamma_1+\gamma_2)/m) = 
-(\gamma_1\gamma_2+mK)(\gamma_1+\gamma_2)/m < 0$ and that $\tilde{g}(x)>0$ 
for any real $x>0$. 
This implies that $\Omega_1 > -(\gamma_1+\gamma_2)/m$ and 
$\text{Re}\{\Omega_{2,3}\}<0$. Consequently, the nonzero roots
$\omega_k=i\Omega_k$~($k=1,2,3$) of $D(\omega)$ are located in the lower 
half-plane. The nonzero roots $\omega_{3+k}=-\omega_k$~($k=1,2,3$) 
of $D(-\omega)$ are located in the upper half-plane~(see
Fig.~\ref{fig3}).

We identified all the roots of $D(\omega)D(-\omega)$. Among them the 
degenerate root $\omega=0$ do not contribute to the integral in
\eqref{residue_int} because it is canceled with $\omega^2$ 
in the numerator of the integrand. The contour in Fig.~\ref{fig3}
includes only $\omega_4$, $\omega_5$, and $\omega_6$. Therefore, the
integral in \eqref{residue_int} is equal to
\begin{equation}
M_{ij} = 2\pi i \left[ \text{Res}(\omega_4) + \text{Res}(\omega_5) +
\text{Res}(\omega_6)\right], 
\end{equation}
where $\text{Res}(\omega_k)$ denotes the residue of the integrand function
in \eqref{residue_int} at the simple pole $\omega_k$. The roots and the
residues are easily evaluated by using a software capable of 
symbolic algebra, which yields the results in \eqref{M2_v}.

We can generalize the analysis to the system with the potential energy
\begin{equation}
V(x_1,x_2) = \frac{k}{2}(x_1^2+x_2^2) + \frac{K}{2}(x_1-x_2)^2
\end{equation}
instead of the potential energy in \eqref{V12}. For simplicity, we set
$\gamma_1=\gamma_2=\gamma$. This system has the 
the modified matrix $\mathsf{F}(\omega)$ whose
elements are $F_{11}(\omega)=F_{22}(\omega) = -m\omega^2-i\gamma \omega+k+K$
and $F_{12}(\omega) = F_{21}(\omega) = -K$. 
Then, $D(\omega) = \det\mathsf{F}(\omega)$ has four roots at
\begin{equation} 
\begin{split}
\omega_{1,2} &= [-i\gamma \pm \sqrt{4mk-\gamma^2}]/({2m}), \\
\omega_{3,4} &= [-i\gamma \pm \sqrt{4m(k+2K)-\gamma^2}]/({2m})
\end{split}
\end{equation} 
in the lower half-plane, and $D(-\omega)$ has four roots at
$\omega_{4+j}=-\omega_{j}$~($j=1,\cdots,4$) in the upper half-plane.
Therefore, the integral corresponding to \eqref{residue_int} 
is determined by the residues at $\omega_{5,6,7,8}$.  
The results are
\begin{equation}
\begin{split}
M_{11} &=  \frac{T_1}{m} - \frac{K^2 (T_1-T_2)}
{2\{mK^2+\gamma^2(k+K)\}}, \\
M_{22} &= \frac{T_2}{m} + \frac{K^2 (T_1-T_2)}
{2\{mK^2+\gamma^2(k+K)\}}, \\
M_{12} &= M_{21} = 0 . 
\end{split}
\end{equation}
Inserting these into Eqs.~\eqref{q1_simple}, \eqref{qS}, \eqref{qF}, and
\eqref{qm}, we obtain that
\begin{equation}\label{kKcase}
\begin{split}
q   & = \frac{\gamma K^2}{2 \{ m K^2 + \gamma^2 (k+K)\}}(T_1-T_2), \\
q_S & = q_F = \frac{\gamma}{4m}(T_1-T_2),\\
q_{m} &= -\frac{\gamma}{2m} 
      \frac{\gamma^2 (k+K)}{\{m K^2 + \gamma^2(k+K)\}}(T_1-T_2) .
\end{split}\end{equation}
In the $K\to\infty$ limit, we obtain that 
$q_m\to0$ and $q_S=q_F\to q_{\infty}/2 = q_{\rm rigid}$.

\begin{acknowledgments}
This work was supported by the Basic Science Research Program through the
NRF Grant No.~2013R1A2A2A05006776.
\end{acknowledgments}

\bibliographystyle{apsrev}
\bibliography{paper}

\end{document}